%% file: main.tex
\documentclass[]{article}

\usepackage[a4paper,total={6in,8in}]{geometry}
\usepackage[square,numbers]{natbib}
\usepackage{amsmath}
\usepackage{amssymb}
\usepackage{amsfonts}
\usepackage{amsthm}
\usepackage{graphicx}
\usepackage{float}
\usepackage{caption}
\usepackage{subcaption}
\usepackage[hidelinks]{hyperref}
\usepackage{authblk}
\usepackage{booktabs}
\usepackage{longtable}
\usepackage{array}

\graphicspath{{figures/}}

\author[1]{Joshua Caiata}
\author[1]{Sreepriya Pulyassary}
\author[1]{Xiang Li}
\author[1]{Kate Larson}
\affil[1]{University of Waterloo}
\affil[ ]{\texttt{jcaiata@uwaterloo.ca, sreepriya.pulyassary@uwaterloo.ca}\\
\texttt{x247li@uwaterloo.ca, kate.larson@uwaterloo.ca}}

\title{Strategy, Not Payoffs: A Behavioural Embedding of Normal-Form Games}
\date{}

\begin{document}

\maketitle

\begin{abstract}
\input{sections/abstract}
\end{abstract}

\input{sections/introductionNEW}

\input{sections/related_work}

\input{sections/preliminaries}

\input{sections/embedding}
\input{sections/method}

\input{sections/experiments}

\input{sections/results}
\input{sections/discussionNEW}
\input{sections/conclusion}

\bibliographystyle{plainnat}
\bibliography{references}

\clearpage
\appendix
\input{sections/appendix}

\end{document}

%% file: sections/abstract.tex
Learning a strategic task changes more than what is directly taught: fine-tuning on one game can either enhance or degrade an agent's ability to reason in another. Understanding and predicting this transfer of strategic capabilities, however, remains a key challenge for large language models (LLMs). Normal-form games provide an ideal testbed for analyzing this phenomenon, as they feature explicitly defined payoffs and well-characterized equilibrium behaviours. In this work, we investigate whether game embeddings can explain and predict changes in LLM strategic capabilities following fine-tuning across different games. We propose a lightweight two-feature embedding that captures fundamental behavioural demands: the entropy of the Nash equilibrium and the sensitivity of optimal responses to an opponent's action. We show that while existing published structural embeddings primarily memorize game identities and fail to generalize, our behavioural embedding reliably predicts performance changes on held-out games. These results demonstrate that the transfer of strategic capabilities in LLMs is not dictated by the payoff geometry of a game, but by the underlying structure of the decision-making behaviour it requires.

%% file: sections/introductionNEW.tex
\section{Introduction}

\begin{figure}[t]
\centering
\includegraphics[width=\textwidth]{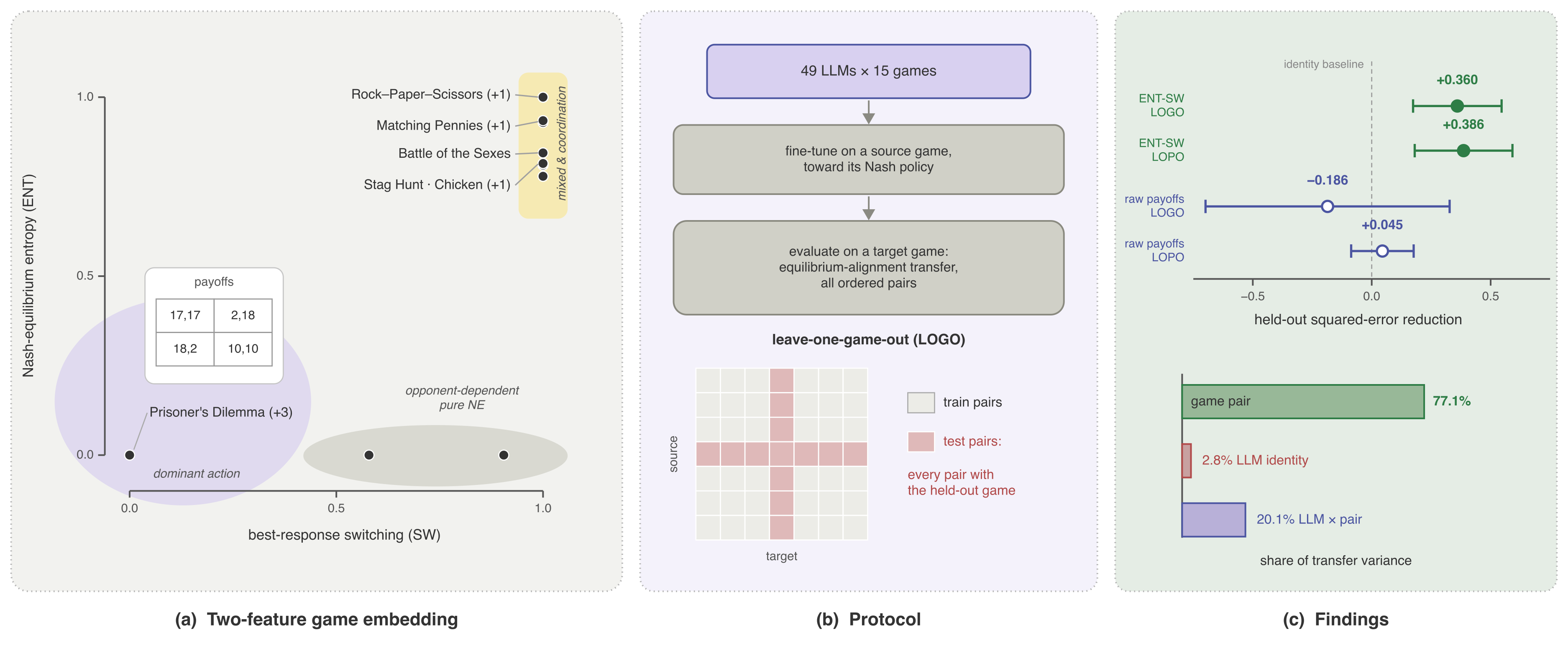}
\caption{
    Overview. (a) 15 games embedded by Nash-equilibrium entropy (ENT) and best-response switching (SW); $(+n)$ marks co-located games. (b) Each of 49 LLMs is fine-tuned toward each source game’s Nash policy and evaluated on all targets. LOGO holds out one game; LOPO one unordered pair. (c) ENT-SW beats game identity in both regimes; RSTP does not. Game-pair identity explains 77.1\% of transfer variance versus 2.8\% for LLM identity.
}
\label{fig:teaser}
\end{figure}

Large language models (LLMs) increasingly act as strategic agents that bargain, cooperate, and compete, often reproducing human-like departures from equilibrium \cite{dafoe2020cooperative,dafoe2021cooperative,park2023generative,camerer2003behavioral,cerapalatsi2025cooperation}. Fine-tuning adapts them to these roles, but training a model on one task may improve or harm its behaviour on another. Canonical normal-form games offer an ideal test case for transfer: they are well-defined, admit computable Nash-equilibrium targets \cite{nash1951noncooperative}, and differ along interpretable dimensions—such as whether equilibrium play is pure or mixed, and how sensitive optimal actions are to opponent behaviour.

Transfer is often predicted using distances in a task-embedding space, as seen in vision and NLP \cite{achille2019task2vec,zamir2018taskonomy,vu2020transferability}. Game theory provides rich material for such embeddings, spanning ordinal taxonomies to equilibrium-invariant and learned representations \cite{rapoport1966taxonomy,marris2023equilibrium,liu2024nfgtransformer}. However, these representations are typically validated in-sample—either by recovering known taxonomies or fitting the very transfer measurements they are scored on \cite{kapoor2022leakage}. Because they are never evaluated on unseen games, a fundamental question remains open: can a game's structure predict behavioural transfer to a genuinely held-out game better than a simple game-identity baseline?

We predict transfer using a two-feature embedding of a game's equilibrium landscape, termed ENT-SW, which captures required behaviour rather than payoff geometry. For each game instance, ENT-SW is computed solely from that instance's payoff table, and is averaged to obtain one representation for each game (Figure~\ref{fig:teaser}a). Equilibrium entropy (ENT) measures ambiguity: near zero for stable pure strategies, and near one when equilibrium requires mixing. Best-response switching (SW) captures sensitivity to opponent behaviour, distinguishing dominant-action games from coordination or cyclic ones. Because neither feature is fit to model behaviour, the embedding avoids data leakage, ensuring every prediction tests a specific structural hypothesis.

We evaluate this embedding using a rigorous generalization protocol (Figure~\ref{fig:teaser}b). We separately fine-tune 49 small open-weight LLMs on every game, using each sampled game instance's Nash equilibrium as its training target. We then measure equilibrium-alignment transfer: the average improvement in matching the equilibria of sampled instances from a target game after fine-tuning on instances from a source game. Transfer predictions are tested under two cross-validation regimes: strict leave-one-game-out (LOGO), testing generalization to an unseen game, and leave-one-pair-out (LOPO), testing interpolation among seen games with unseen pairings. ENT-SW is benchmarked against a one-hot game-identity baseline. To prevent correlated trials from inflating fit, each LLM-source-target cell is averaged over trials, and per-LLM fixed effects absorb model-level variations. Figure~\ref{fig:teaser}c provides an overview of the results.

Under this protocol, ENT-SW successfully predicts transfer to held-out games. It outperforms the game-identity baseline in both LOGO and LOPO settings, retaining significant predictive value even after identity receives full credit (see Results). In contrast, structural embeddings from the literature only beat the identity baseline under LOPO interpolation; their advantage vanishes on genuinely new games. This demonstrates that predicting transfer requires representing the behaviour a game demands, rather than merely the geometry of its payoffs.
Our architecturally diverse roster also clarifies whether transfer is an artifact of the specific model fine-tuned. Decomposing transfer variance reveals that game-pair structure accounts for 77.1\% and the LLM-pair component explains 20.1\%, while model identity explains only 2.8\%. Within pairs, the source-target interaction dominates the individual contributions of either game. Transferability is thus fundamentally a property of the game pair, a property our embedding makes predictable.



%% file: sections/related_work.tex
\section{Related Work}

\textbf{Predicting transfer from task representations.}
A common paradigm forecasts transferability by embedding tasks as vectors and measuring distances within that space. For example, Task2Vec derives fixed-length vectors from a probe network's Fisher information \cite{achille2019task2vec}; Taskonomy estimates a pairwise transfer-affinity matrix across vision tasks \cite{zamir2018taskonomy}; and NLP task embeddings built from fine-tuning gradients rank source tasks by similarity \cite{vu2020transferability}. Lighter-weight scores like LEEP \cite{nguyen2020leep}, LogME \cite{you2021logme}, and dataset distances like OTDD \cite{alvarezmelis2020otdd} forecast transfer from a single forward pass or dataset comparison, while recent work explores low-dimensional, visualizable representations \cite{daroya2024task2box}. Typically, these representations are learned, opaque, or evaluated in-sample—against their training transfer matrix or by top-$k$ source-selection regret—rather than on held-out, unseen tasks. Moreover, measured transfer varies sharply across sources and seeds \cite{lin2024consistency}, making metric evaluations unstable under varying experimental settings \cite{agostinelli2022stable}. Current methods focus on vision and NLP classification; none benchmark against a task-identity control or separate portable structure from model-identity variation. We import this embedding-based transfer paradigm into strategic games. However, we replace the opaque learned vector with a two-feature embedding of a game's equilibrium landscape, evaluated under a falsifiable leave-one-game-out protocol against an explicit game-identity baseline.

\textbf{LLMs as strategic agents.}
A growing literature evaluates large language models in canonical strategic games. Most studies prompt models to play social dilemmas or coordination games and compare their choices to rational or Nash play. This work documents systematic biases \cite{akata2023repeated,fan2023rational}, builds benchmarks \cite{duan2024gtbench}, and improves sub-optimal play via prompting scaffolds, solvers, or iterative reasoning \cite{gandhi2023strategic,gemp2024steering,kempinski2025gameofthoughts}. Consistently, models exhibit persistent strategic fingerprints across games \cite{payne2025strategic,sun2025survey}. Shifting from prompting to fine-tuning, \citet{lore2024strategic} train a single small model to imitate a larger model's responses across several social dilemmas. However, this literature remains largely evaluative, covers few games, and lacks an equilibrium-alignment target. Even the fine-tuning study distills teacher imitation rather than a Nash distribution, evaluates a single model, and fails to separate game structure from model identity. We instead separately fine-tune 49 small open-weight LLMs on every game using instance-specific Nash-equilibrium targets. We then predict cross-game equilibrium-alignment transfer using an interpretable structural embedding, explicitly isolating this structural transfer from the model-identity variance caused by per-model strategic fingerprints.


\textbf{Structural representations of strategic games.}
A parallel line of work develops structured game representations. Classical game theory organizes $2\times2$ games by ordinal payoff structure \cite{rapoport1966taxonomy,robinson2005topology}. Recent work extends this by formalizing equilibrium-invariant embeddings of normal-form games \cite{marris2023equilibrium} and learning permutation-equivariant representations via neural networks \cite{liu2024nfgtransformer}. Complementarily, empirical game-theoretic analysis builds reduced ``meta-game'' models from agent simulations to make large strategic spaces tractable \cite{wellman2024egta,omidshafiei2019alpharank, Omidshafiei2020}. However, these representations are largely validated in-sample—either by recovering a known taxonomy or solving equilibria on the training game—rather than as predictors of cross-game behavioural transfer. We empirically close this gap. We compute representative embeddings from prior work, including equilibrium- and better-response-invariant embeddings \cite{marris2023equilibrium}, response-graph statistics \cite{omidshafiei2019alpharank}, and potential-harmonic decompositions \cite{candogan2011flows}. We evaluate each as a baseline under our held-out protocol to test whether they successfully predict behavioural transfer to unseen games.

%% file: sections/preliminaries.tex
\section{Preliminaries}
\label{sec:prelim}

\subsection{Normal-Form Games}

We consider finite, two-player normal-form games. We use $\mathcal{G} = \{G_1,\ldots,G_C\}$ to denote our set of games, where $C=15$. A game $G_i \in \mathcal{G}$ denotes a named strategic setting, such as Prisoner's Dilemma or Stag Hunt. A game \emph{instance}, $G_i^k$ is a normal-form payoff table following some constraints that satisfy $G_i$. $G_i^k$ is defined by a tuple, $(P, \{\mathcal{A}_p\}_{p \in P}, \{u^k_p\}_{p \in P})$, where $P = \{1,2\}$ is the set of player roles, $\mathcal{A}_p$ is the finite action set available to player $p$, and $u^k_p : \mathcal{A}_1 \times \mathcal{A}_2 \rightarrow \mathbb{R}$ is player $p$'s payoff function. For player $p$, action $a \in \mathcal{A}_p$, and opponent action $b \in \mathcal{A}_{p^{-1}}$ (where $p^{-1}$ denotes $p$'s opponent), $u_p(a,b)$ gives the payoff to player $p$ when the two players play $a$ and $b$ respectively. Most games we consider are 2-action, 2-player games, except for Rock-Paper-Scissors and Bertrand Duopoly, which are 3-action, 2-player. A game is \emph{symmetric} if $\mathcal{A}_1 = \mathcal{A}_2$ and $u_1(a,b) = u_2(b,a)$ for all $a,b$; otherwise it is \emph{asymmetric}, and the two player roles may face qualitatively different strategic problems.

\subsection{Strategies and Best Responses}

A strategy for player $p$ is a probability distribution $\pi_p$ over $\mathcal{A}_p$, with $\pi_p(a)$ denoting the probability assigned to action $a$. Given an opponent action $b \in \mathcal{A}_{p^{-1}}$, player $p$'s set of pure best responses is $BR_p(b) = \operatorname*{arg\,max}_{a \in \mathcal{A}_p} u_p(a,b).$ When $|BR_p(b)| > 1$, player $p$ is indifferent among multiple actions against $b$.

\subsection{Nash Equilibrium}

A strategy profile $(\pi_1, \pi_2)$ is a Nash equilibrium if each player's strategy is a best response to the other's: $\operatorname{supp}(\pi_p) \subseteq BR_p(\pi_{p^{-1}})$, for $p \in \{1,2\}$, where $BR_p(\pi_{p^{-1}})$ is the set of actions maximizing player $p$'s expected payoff against the mixed strategy $\pi_{p^{-1}}$, and $\operatorname{supp}(\cdot)$ denotes the support of a distribution. Every finite normal-form game admits at least one Nash equilibrium, possibly in mixed strategies \citep{nash1951noncooperative}. Some games in our suite admit a unique pure-strategy equilibrium, while others require mixed strategies; several admit multiple equilibria, in which case we adopt the mixed equilibrium described in Appendix~\ref{ap:games}. For a given game instance $G_i^k$ and player role $p$, we write $\pi_{G_i^k,p}$ for the (selected) Nash equilibrium strategy, which serves as the behavioural target used throughout our evaluation.

\subsection{Behavioural Evaluation Setup}
Rather than evaluating whether a model can compute a Nash equilibrium symbolically, we treat $\pi_{G_i^k,p}$ as a target \emph{behavioural policy}: given a description of a game instance, we ask whether the language model's induced action distribution matches the equilibrium distribution over actions. This framing lets us pose game play as a distributional prediction problem, and lets us later ask how equilibrium behaviour learned in one game (a source game) transfers to another (a target game), which we formalize in the next section.

%% file: sections/embedding.tex
\section{Strategic Embedding of Normal-Form Games}

We now detail our proposed embedding, which we call ENT-SW (which simply refers to entropy, as in Nash entropy, and switching, as in best-response switching). Specifically, the primary structural embedding we propose in this paper is a two-dimensional hand-crafted feature vector: $h(G_i^k, p) = [H_{NE}(G_i^k,p), S_{BR}(G_i^k,p)]$, where $H_{NE}(G_i^k,p)$ is the normalized entropy of the Nash equilibrium policy for player $p$. Concretely, if $\pi_{G_i^k,p}(a)$ is the Nash probability assigned to action $a$, then:

\[
H_{NE}(G_i^k,p) = -\frac{1}{\log |\mathcal{A}_p|} \sum_{a \in \mathcal{A}} \pi_{G_i^k,p}(a) \log \pi_{G_i^k,p}(a)
\]

Pure-equilibrium games have entropy near 0, while games requiring mixing have entropy near 1.

The second feature, $S_{BR}(G_i^k,p)$, measures best-response switching. This captures whether the same action is optimal against every opponent action, or whether the player's best response depends on what the opponent does. We then compute the empirical distribution of best-response actions across opponent actions:

\[
q_{G_i^k,p}(a) = \frac{1}{|\mathcal{A}_{p^{-1}}|} \sum_{b \in \mathcal{A}_{p^{-1}}} \frac{\mathbf{1}[a \in BR_p(b)]}{|BR_p(b)|}
\]

The best response feature is then the normalized entropy of this distribution:

\[
S_{BR}(G_i^k,p) = -\frac{1}{\log |\mathcal{A}_p|} \sum_{a \in \mathcal{A}_p} q_{G_i^k,p}(a) \log q_{G_i^k,p}(a)
\]

An $S_{BR}(G_i^k,p)$ value near 0 means the same action is always a best response, where a value near 1 indicates that best responses switch across opponent actions (as seen in coordination, anti-coordination, or cyclic games). 

Because we measure transfer between games as opposed to individual game instances (as in, the Prisoner's Dilemma as a whole and how it might transfer to Stag Hunt as a whole), we need to construct one aggregate representation for each game. To do so, we use a canonical per-game vector obtained by averaging the embedding over the sampled off-diagonal transfer instances and relevant player roles, computed as $\bar{h}(G_i) = \frac{1}{N_i} \sum_{k=1}^{N_i} h(G_i^k, p_k)$. To see a visualization of the embedding of the games on a scatterplot, refer to Figure \ref{fig:teaser}a.

For a directed transfer pair between source game $G_i$ and target game $G_j$, we construct the pair of representations as: $\phi(i,j) = [\bar{h}(G_i), \bar{h}(G_j), |\bar{h}(G_i) - \bar{h}(G_j)|, \bar{h}(G_i) \odot \bar{h}(G_j)]$ where $\odot$ denotes element-wise multiplication.

%% file: sections/method.tex
\section{Transfer Evaluation Setup}
\label{sec:methods}

\subsection{Game Suite and Equilibrium Targets}

We select a wide array of classic, well-studied games to evaluate behavioural transfer. We use 15 games in our work: Prisoner's Dilemma, Stag Hunt, Harmony, Chicken, Battle of the Sexes, Matching Pennies, Entry Game, Bertrand Duopoly, Rock-Paper-Scissors, Deadlock, Compromise, Pure Coordination, Volunteer's Dilemma, Trust Game, and Inspection Game. For asymmetric games, the Nash equilibrium is computed from the relevant player's role. For full details about the games and their Nash equilibria, refer to Appendix \ref{ap:games}. We do not propose that this is an exhaustive list of games, however, we aimed to select a wide variety from the literature. 

Across all games, the equilibrium target used for evaluation is, therefore, a distribution over the available actions, not just a single best action. If a game admits multiple Nash equilibria, we select the mixed Nash equilibrium. During evaluation, the model produces a probability distribution over the randomized action labels in the prompt. The target Nash distribution is mapped onto those same labels, and behavioural error is measured as mean squared error between the model distribution and the Nash equilibrium distribution, defined as: $\mathrm{MSE} = \frac{1}{|\mathcal A_p|} \sum_{a \in \mathcal A} \left( P_{\mathrm{model}}(a) - P_{\mathrm{NE}}(a) \right)^2$.

\subsection{Data Generation}
For each game, we generate a supervised fine-tuning (SFT) dataset by sampling payoff instances, given each game's payoff constraints. For each instance, we compute the Nash equilibrium for the relevant player roles. Symmetric games use a fixed player perspective, while asymmetric games sample a role based on which the equilibrium target is computed.

The prompts describe only the payoff consequences of the available actions and do not reveal the game name. Action labels are randomized uppercase letters, and the action order is also randomized, so the model cannot rely on fixed labels or positions. The Nash distribution over actions is then mapped onto the randomized letters shown in the prompt.

The output label for each training example is sampled from this mapped Nash distribution; each training example includes a prompt, a sampled output letter, an action-label mapping, and the distribution of the Nash equilibrium for that agent. This yields one dataset per game, which is used to fine-tune the source-game model.

\subsection{Fine-Tuning}
We fine-tune models separately for each game. For each base LLM $m$ and source game $G_i$, we train a distinct supervised fine-tuned model on the dataset generated from $G_i$. Thus, each fine-tuned model specialized in one source game rather than being trained jointly across the full game suite.

Each training example consists of a payoff prompt and a sampled Nash-equilibrium action label. During SFT, the prompt tokens are masked from the loss, so the model is trained only to predict the answer action. This encourages the model to learn the behavioural policy associated with the source game's equilibrium target.

After fine-tuning, the resulting checkpoint is treated as the source-game model for transfer evaluation. For a suite with $M$ base LLMs and $C$ games, this gives one fine-tuned model for each pair $(m, G_i)$, producing $M \times C$ source-specialized models. The models are then evaluated on held-out target games to measure behavioural transfer.

\subsection{Behavioural Transfer Metric}
We measure behavioural transfer as improvement over the corresponding cold base model on the same target game. For a target game $G_j$, the cold error is the average NE-MSE of the base LLM $m$ before game-specific fine-tuning: $\mathrm{MSE}_{\mathrm{cold}}(m,j) = \frac{1}{K}\sum_{k=1}^{K} \mathrm{MSE}_{m,j,k}^{\mathrm{cold}}$. Then, for an ordered source-target pair $(G_i,G_j)$, the tuned error is the average NE-MSE when the model fine-tuned on source game $G_i$ is evaluated on target game $G_j$: $\mathrm{MSE}_{\mathrm{tuned}}(m,i,j) = \frac{1}{K}\sum_{k=1}^{K} \mathrm{MSE}_{m,i,j,k}^{\mathrm{tuned}}$.

We define behavioural transfer using a log-relative improvement score:

\[
\mathrm{LTS}(m,i,j)
=
\log\left(
\frac{
\mathrm{MSE}_{\mathrm{cold}}(m,j)+c
}{
\mathrm{MSE}_{\mathrm{tuned}}(m,i,j)+c
}
\right)
\]

where $c>0$ is a small offset used to keep the log well-defined and stabilize very small error values, which we compute using the median standard error over our Monte Carlo trials, and fix $c \approx 0.003$. Positive values indicate beneficial transfer (i.e., the source-game fine-tuned model has lower target-game error than the cold base model), while values near zero indicate little change.

\subsection{Pooled Cross-LLM Framework}
We pool evaluation results into cells indexed by base LLM $m$, source game $G_i$, and target game $G_j$. The response variable is then the transfer score, $LTS(m,i,j)$, where we only use the off-diagonal transfer cells (i.e. we exclude instances where $i = j$.) For each feature representation $A$, we fit a weighted ridge regression of the form $LTS(m,i,j) = \beta_0 + \alpha_m + \epsilon_{mij} + \beta_A^\top \phi_A(i,j)$, where $\alpha_m$ is the intercept for the LLM, encoded using $M-1$ indicators, $\phi_A(i,j)$ is the feature vector, $\beta_0$ is the whole-model intercept, $\beta_A$ are the learned coefficients, and $\epsilon_{mij}$ is the cell-specific residual. The loss function has a residual defined as: $r_{mij} = LTS(m,i,j)-\beta_0-\alpha_m-\beta_A^\top\phi_A(i,j)$, and the fitted objective is therefore: $\min_{\beta_0,\alpha,\beta_A} \sum_{m,i,j} w_m r_{mij}^2 + \lambda \lVert \beta_A\rVert_2^2$, where $w_m$ is a family-balanced weight, where each model is weighted inversely by the number of evaluated models in its architecture family, preventing overrepresented families from dominating the pooled regression. Continuous embedding features are standardized within each training fold using z-score normalization. $\lambda = 1000$ was selected from a fixed grid based on LOGO performance and then reused across experiments for consistency.

%% file: sections/experiments.tex
\section{Experiments}

\subsection{Experimental Setup}
The production sweep uses a broad roster of small open-weight base LLMs, all at or below 3B parameters. The roster spans major architecture families including Pythia, SmolLM2, GPT-2, GPT-Neo, OPT, Cerebras-GPT, Qwen2.5, OLMo/OLMo-2, TinyLlama, Danube, Phi, StableLM, Falcon, BLOOM, Mamba, Llama-3.2, Gemma, and MobileLLM, totalling 49 LLMs.

For each LLM and game pair, we run $K$ Monte Carlo evaluation trials. In each trial, a fresh target game instance is sampled, action labels are randomized, and the model's action distribution is compared to the Nash equilibrium target. We use $K=2000$ trials per cell, and a seed of 42.

\noindent{\bf Fine-Tuning Runs:}
We separately fine-tune each base LLM for every source game, yielding one checkpoint per LLM-game pair. Models undergo full-model supervised fine-tuning for 5 epochs on 20,000 generated examples. The loss masks prompt tokens, optimizing solely to predict the sampled Nash-equilibrium action label following the response separator.

\noindent{\bf Transfer Evaluation Runs:} After fine-tuning, each LLM is evaluated on a full 15-game source-target grid. This yields 225 tuned cells per LLM: 15 diagonal cells (where source equals target) to verify the model learned its source game, and 210 off-diagonal transfer cells for the main analysis. To establish a cold baseline, each LLM is also evaluated on all 15 target games prior to fine-tuning. With 2000 trials (distinct payoff instances) across each of the 15 cold and 225 tuned cells, we run 480,000 trials per LLM.

During evaluation, models are prompted exclusively with the target game; source-game information enters solely through the checkpoint (the model tuned on source game $G_i$ is evaluated on target game $G_j$). Finally, we record per-trial MSEs and aggregate them into average errors for each LLM-source-target cell.




\subsection{Statistical Test Suite}
We run a test suite on pooled LLM-source-target cells using the log transfer score as our outcome variable. The main comparisons are to test whether an embedding alone predicts transfer better than or as well as game identity, and to test whether adding the embedding to game identity improves prediction over identity alone. 

We evaluate generalization using two types of game-structured cross-validation. Specifically, strict leave-one-game-out (LOGO) where we leave one game out of the set, including any unordered pair $\{i,j\}$ that includes that game. This tests whether the model can predict an unseen transfer relationship on a game it has never seen before. Leave-one-pair-out (LOPO) holds out both directions of an unordered game pair $\{i,j\}$, testing whether the model can predict an unseen transfer when both games have been observed elsewhere. We estimate uncertainty using a size-stratified cluster bootstrap over model families and held-out games. Paired differences in held-out squared error are aggregated by family, size band, and game; families and games are then resampled with replacement, with size bands (<200M, 200M–500M, 500M–1B, 1–2B, and 2–3B) averaged equally. We report 95\% percentile intervals from 10,000 valid replicates. Family-level clustering separately accounts for dependence among models from the same lineage. 

We compare our embedding against several baselines to situate ENT-SW against other analytic structural representations under a common protocol. The primary baseline consists of one-hot indicators for the source and target game: $\phi_{ID} = [e_i, e_j]$. This baseline allows us to test whether an analytic representation can substitute for game-specific labels when predicting transfer, conditional on LLM fixed effects. Under the same evaluation protocol, we compare against non-learned baselines derived from prior work: component-magnitude and component-coordinate features from the potential–harmonic decomposition (PHD energy and PHD proj) \citep{candogan2011flows}, response-graph statistics (RG stats) \citep{Omidshafiei2020}, and equilibrium- and better-response-invariant representations (EI-diff, EIE cloud, and BRI cloud) \citep{marris2023equilibrium}. EI-diff operates on the full payoff matrix, whereas the cloud baselines aggregate all \(2\times2\) subgames. We also include an RSTP raw-payoff baseline for two-action games, using the row player’s payoff vector when cooperation–defection semantics do not apply (Appendix \ref{ap:games}).

Finally, we compute ICC-style variance decompositions \cite{shrout1979intraclass,nakagawa2013r2} to estimate how much transfer is attributable to LLM identity, game-pair structure, and their interactions.

%% file: sections/results.tex
\section{Results}
\label{sec:results}
\begin{figure}[t]
    \centering
    \begin{subfigure}[t]{0.48\textwidth}
        \centering
        \includegraphics[width=0.9\linewidth]{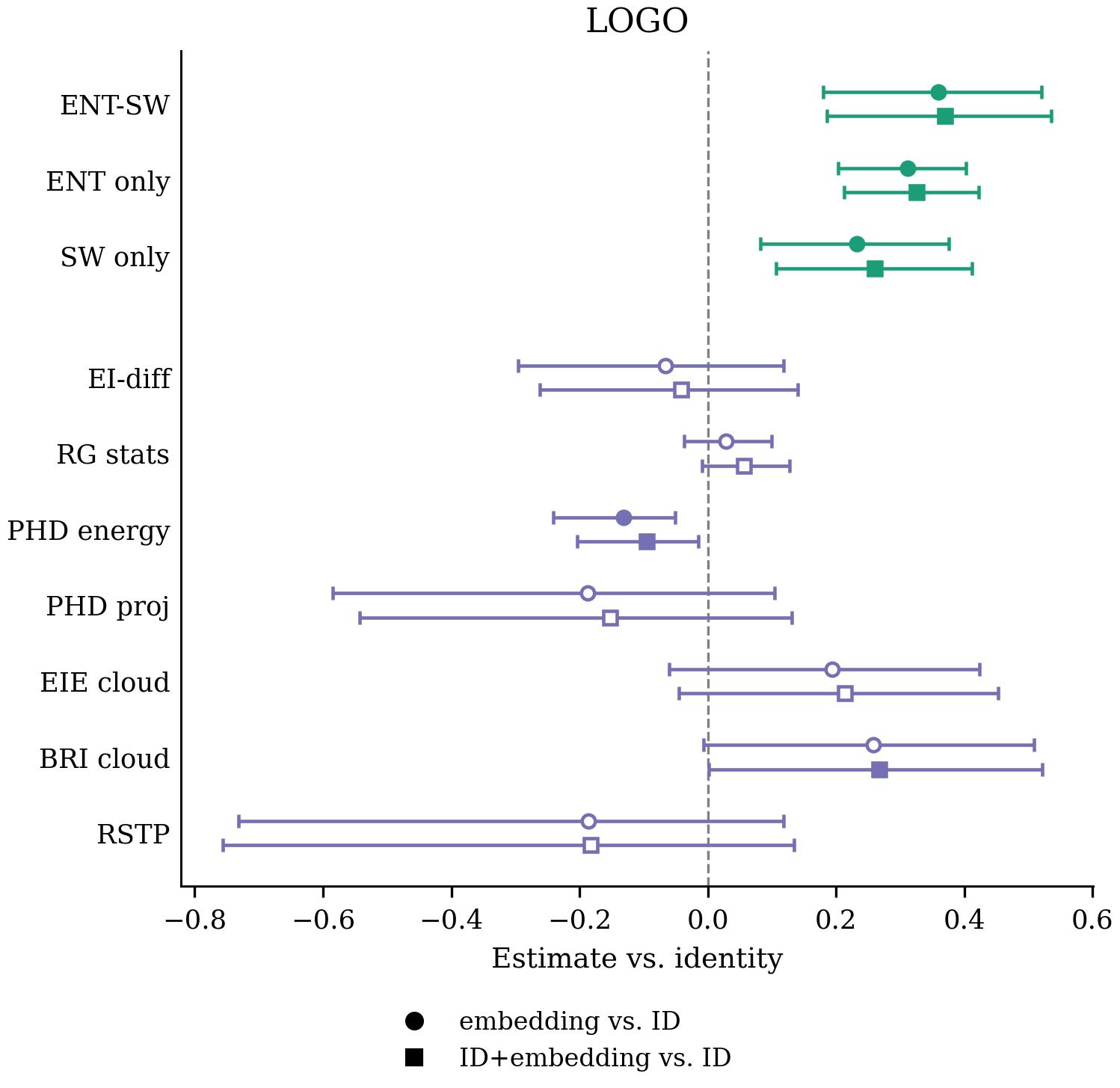}
        \caption{Leave-one-game-out (LOGO)}
        \label{fig:logo}
    \end{subfigure}
    \hfill
    \begin{subfigure}[t]{0.48\textwidth}
        \centering
        \includegraphics[width=0.9\linewidth]{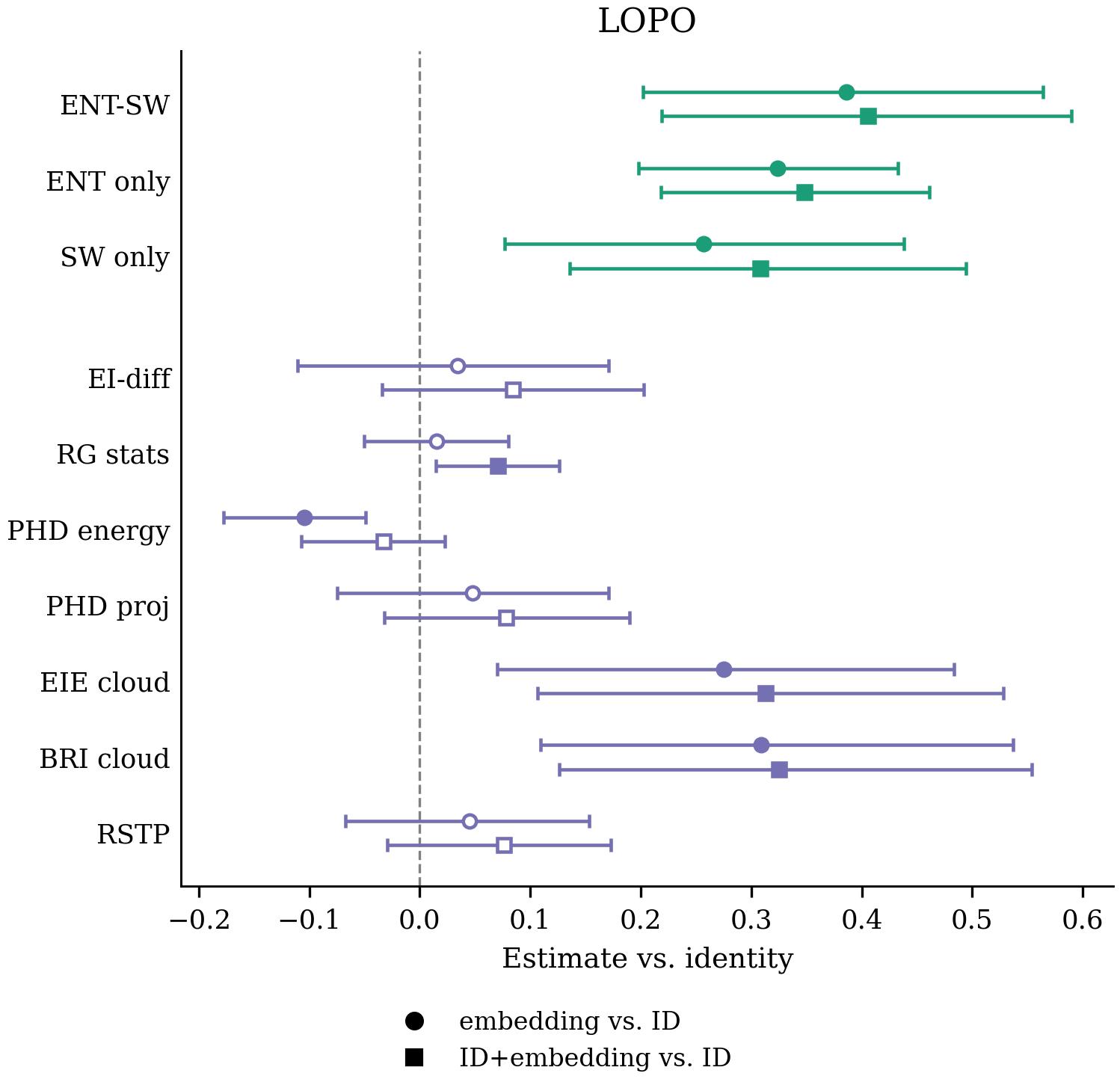}
        \caption{Leave-one-pair-out (LOPO)}
        \label{fig:lopo}
    \end{subfigure}
    \caption{Held-out squared-error reduction relative to game identity under (a) LOGO and (b) LOPO. Positive values favor the embedding. Circles compare each embedding with identity; squares compare identity plus the embedding with identity alone. Bars show 95\% confidence intervals; filled markers exclude 0.
}
    \label{fig:logo-lopo}
\end{figure}

We first directly answer three research questions. For an overview of the results, refer to Figure \ref{fig:logo-lopo}.
 
\noindent{\bf Do Game-Structure Embeddings Predict Transfer as Well as Game Identity?}
ENT-SW outperformed identity by $0.360$ (95\% CI $[0.180, 0.520]$) under strict leave-one-game-out and $0.386$ ($[0.202, 0.564]$) under leave-one-pair-out, with the single-feature ablations ENT only and SW only showing the same pattern. Under strict leave-one-game-out, the held-out game never appeared in any training pairing, so identity had no memorized parameter for it either.

\noindent{\bf Do Game-Structure Embeddings Add Predictive Value Beyond Game Identity?}
When ENT-SW was added to the penalized additive identity specification, it further reduced held-out prediction error, at $0.370$ $[0.185,0.536]$ under strict leave-one-game-out and $0.406$ $[0.219,0.590]$ under leave-one-pair-out, with ENT only and SW only again consistent.

\noindent{\bf How Much Transfer Variation Is Attributable to Game-Pair Structure Versus LLM Identity?}
Game-pair identity explains $77.1\%$ ($[64.3\%,86.3\%]$) of coarse-level variance in transfer scores, against $2.8\%$ ($[0.9\%,8.1\%]$) for LLM identity. Within game pairs, the source$\times$target interaction accounts for $63.5\%$ ($[18.3\%,72.5\%]$) of variance, against $21.5\%$ for source alone and $15.0\%$ for target alone. Model size contributes a smaller, largely separate share ($7.4\%$ main-effect / $26.6\%$ within-family, depending on conditioning, with only $1.6\%$ interaction with pair identity).

\subsection{Is This Specific to the Proposed Embedding?}

\textbf{RSTP, EI-diff, and PHD energy fail outright.} Raw payoff values (RSTP) never distinguish themselves from identity, and in fact point in the wrong direction under strict LOGO. EI-diff, computed on the full payoff matrix, likewise crosses zero in all four comparisons. PHD energy is significantly \emph{worse} than identity under both LOGO and LOPO ($-0.132$ and $-0.104$ respectively); this tells us that a potential–harmonic summary of the game's flow structure, on its own, actively hurts prediction relative to just knowing which game is which.
 
\textbf{RG stats, EIE cloud, and BRI cloud are regime-dependent.} These three features look promising under LOPO; BRI cloud, for example, reaches $0.309$ $[0.110, 0.537]$ under RQ1 pair, yet each of them crosses zero under strict LOGO. LOPO allows the held-out game to have appeared in \emph{other} training pairings, so a feature that partially memorizes pairwise regularities can still look predictive. Under strict LOGO, the held-out game never appears in training at all, and under that harder test these three baselines collapse toward identity-level performance or worse.
 
\textbf{ENT-SW and its ablations are the only features that survive strict LOGO.} ENT-SW, as well as ENT and SW on their own, all exclude zero in their confidence intervals for both research questions under both LOGO and LOPO---the only features in our comparison set with that property. Any equilibrium property is not sufficient; RSTP, also derived directly from the payoff matrix, fails, and the literature baselines are themselves structural embeddings; none replicate ENT-SW's robustness.

\section{Games in our Embedding Space}

\begin{figure}[H]
    \centering
    \includegraphics[width=0.9\linewidth]{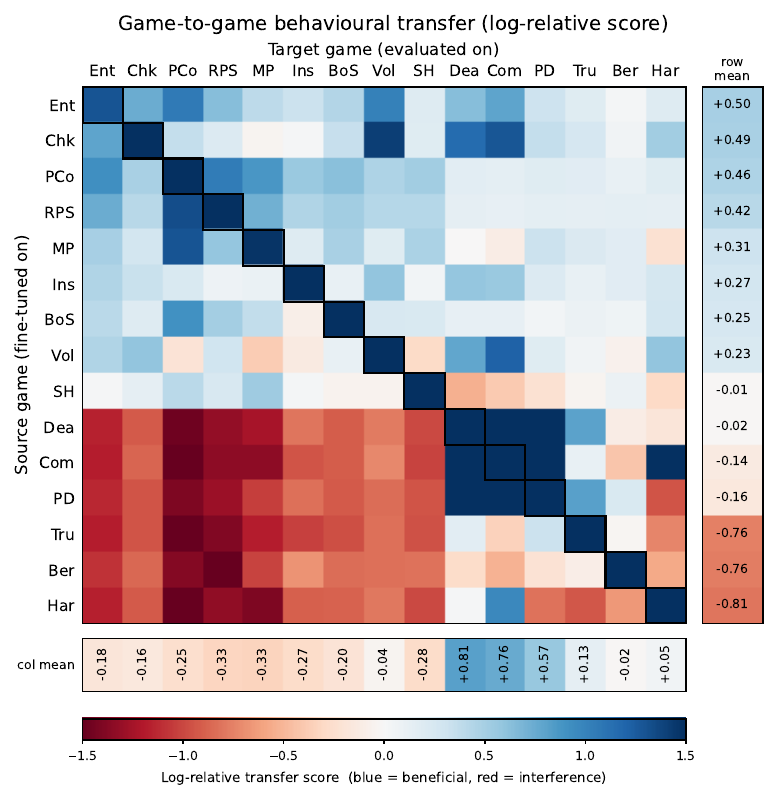}
    \caption{Mean log-relative transfer across LLMs. Rows are source games and columns are targets, sorted by source mean. Blue indicates beneficial transfer; red indicates interference. Game abbreviations correspond to the suite in Appendix~\ref{ap:games}.}
    \label{fig:heatmap}
\end{figure}

Naming the games exposes a limit that the scalar embedding cannot see. Chicken, Entry Game, and Pure Coordination are the strongest general-purpose sources in the suite; fine-tuning on any of them improves performance on the average target, while Harmony, Trust Game, and Bertrand Duopoly are the weakest. Prisoner's Dilemma, Deadlock, and Compromise form the single most compatible cluster in the entire matrix: each transfers into the other two at among the highest rates observed anywhere in the suite. All three share a dominant strategy that resolves toward the same kind of outcome, and $H_{NE}=0$, $S_{BR}=0$ identifies them as behaviorally identical in shape. Figure \ref{fig:heatmap} shows specific transfer details aggregated over every LLM in more detail.

Harmony is the outlier that shows what shape alone cannot capture. It has the same $H_{NE}=0$ and $S_{BR}=0$ as Prisoner's Dilemma, Deadlock, and Compromise, yet it is both a significantly worse source than those three games, and there is no meaningful transfer to or from Harmony and those three games. The difference is that Harmony's dominant strategy resolves toward the cooperative action, whereas the other three resolve toward the self-interested one. A policy fine-tuned to commit unconditionally to self-interest and a policy fine-tuned to commit unconditionally to cooperation are both maximally confident, zero-entropy, zero-switching policies by the embedding's own measure, and almost perfect opposites behaviorally. $H_{NE}$ and $S_{BR}$ describe the shape of an equilibrium without reference to which action it points toward, so two games can be indistinguishable in our embedding space while being about as behaviorally incompatible as this suite contains. Despite this property, however, it is still the best in our comparison set at predicting transfer.

This is consistent with the variance decomposition: if transfer is governed by how compatible two games' equilibrium landscapes are, the outcome should be dominated by the specific source-target relationship rather than by either game's difficulty in isolation or by which model is transferring. This is clearly seen by the source$\times$target interaction accounting for the majority of within-pair variance and LLM identity contributing comparatively little.

%% file: sections/discussionNEW.tex
\section{Discussion}
A strategic game embedding is only as valuable as the behaviour it captures. Predicting transfer directly tests whether an embedding captures a game's behaviourally relevant structure, rather than a coincidental surface statistic.

Game-pair identity serves as the ideal baseline because it captures the average transferability of each observed source game and the average receptivity of each observed target game. Such a flexible baseline should only lose to a two-dimensional embedding if the embedding encodes generalizable information that identity inherently cannot—specifically, information about pairings unseen in training. Because ENT-SW outperforms identity under strict leave-one-game-out (where the held-out game is completely unseen), it clearly captures this generalizable structure. Furthermore, ENT-SW's significant marginal contribution, even when identity's variance is fixed, proves it is not merely reconstructing identity implicitly.

RSTP, PHD energy, RG statistics, and the BRI-family features fail to exhibit this pattern. RSTP, PHD energy, and RG statistics never beat the identity baseline in either regime. While the BRI-family features outperform identity under leave-one-pair-out—where both games in a held-out pair may have been observed elsewhere in training—this advantage vanishes under the stricter leave-one-game-out test. This highlights the difference between features that merely interpolate among familiar games (which identity already accomplishes) and those that generalize to genuinely new games. ENT-SW is the only tested representation that meets this strict generalization criterion.

$H_{NE}$ and $S_{BR}$ share an underlying intuition. Equilibrium entropy measures strategic ambiguity: a game with a sharply dominant equilibrium dictates stable behaviour, whereas a diffuse or multi-modal equilibrium requires a policy to select or maintain a position within that ambiguity. Best-response switching complements this by describing dynamics rather than destination; specifically, how optimal play shifts in response to an opponent's behaviour. This dictates how reactive or stable a learned policy must be to remain near equilibrium. Together, these two quantities describe the strategic terrain a policy navigates and its required responsiveness.

This aligns with our variance decomposition. If transfer depends on the compatibility of two games' equilibrium landscapes, the specific source-target relationship should dominate the outcome, rather than individual game difficulty or transferring model identity. We observe exactly this pattern: the source$\times$target interaction accounts for the vast majority of within-pair variance, while LLM identity contributes comparatively little.

%% file: sections/conclusion.tex
\section{Conclusion}

We asked whether normal-form games can be represented by a compact embedding that meaningfully explains the behaviour they induce in LLMs when used for fine-tuning. A two-feature embedding built from Nash-equilibrium entropy and best-response switching outperformed a fully flexible game-identity baseline and remained the only structural representation we tested to do so under strict leave-one-game-out evaluation. Because identity already explains most of the variation in transfer outcomes, the practical value of this embedding lies in not only outperforming identity outright, but also in offering a compact, generalizable stand-in when identity cannot be used at all: prediction for a game the model has never encountered.

Our results do not establish that equilibrium entropy and best-response switching are uniquely responsible for this predictive power; a differently constructed pair of equilibrium-based features might perform comparably or better. Our suite is also limited to fifteen two- and three-action games, and whether these features continue to predict transfer in games with larger action spaces, more than two players, or sequential structure remains untested.

More broadly, these results suggest that behavioural transfer is better predicted by the shape of a game's equilibrium landscape than by its raw payoff geometry. Extending this approach to richer games is a natural direction for future work.

%% file: sections/appendix.tex
\section{Game Suite and Nash Equilibrium Targets}
\label{ap:games}

\setlength{\tabcolsep}{4pt}
\renewcommand{\arraystretch}{1.75}

\begin{longtable}{
  >{\raggedright\arraybackslash}p{0.15\textwidth}
  >{\raggedright\arraybackslash}p{0.34\textwidth}
  >{\raggedright\arraybackslash}p{0.20\textwidth}
  >{\raggedright\arraybackslash}p{0.23\textwidth}
}
\caption{Game suite and Nash equilibrium targets used in evaluation. Each payoff cell is written as (row payoff, column payoff). Action labels shown to models are randomized, but the table reports the underlying action indices.}\\
\toprule
\textbf{Game} & \textbf{Payoff matrix / actions} & \textbf{Requirements} & \textbf{Equilibrium target} \\
\midrule
\endfirsthead

\toprule
\textbf{Game} & \textbf{Payoff matrix / actions} & \textbf{Requirements} & \textbf{Equilibrium target} \\
\midrule
\endhead

Prisoner's Dilemma (PD)
&
$\begin{pmatrix}
(R,R) & (S,T) \\
(T,S) & (P,P)
\end{pmatrix}$

Action 0: cooperate; action 1: defect.
&
$T>R>P>S$; additionally $2R>T+S$.
&
Pure defection:
$\pi(0)=0,\ \pi(1)=1$.
\\
\midrule

Stag Hunt (SH)
&
$\begin{pmatrix}
(R,R) & (S,T) \\
(T,S) & (P,P)
\end{pmatrix}$

Action 0: cooperate; action 1: defect.
&
$R>T>P>S$; additionally $2R>T+S$.
&
Mixed target:
$\pi(0)=\frac{P-S}{R-S-T+P}$,
$\pi(1)=1-\pi(0)$.
\\
\midrule

Chicken (Chk)
&
$\begin{pmatrix}
(R,R) & (S,T) \\
(T,S) & (P,P)
\end{pmatrix}$

Action 0: cooperate; action 1: defect.
&
$T>R>S>P$.
&
Mixed target:
$\pi(0)=\frac{P-S}{R-S-T+P}$,
$\pi(1)=1-\pi(0)$.
\\
\midrule

Harmony (Har)
&
$\begin{pmatrix}
(R,R) & (S,T) \\
(T,S) & (P,P)
\end{pmatrix}$

Action 0: cooperate; action 1: defect.
&
$R>T>S>P$.
&
Pure cooperation:
$\pi(0)=1,\ \pi(1)=0$.
\\
\midrule

Deadlock (Dea)
&
$\begin{pmatrix}
(R,R) & (S,T) \\
(T,S) & (P,P)
\end{pmatrix}$

Action 0: cooperate; action 1: defect.
&
$T>P>R>S$.
&
Pure defection:
$\pi(0)=0,\ \pi(1)=1$.
\\
\midrule

Compromise (Com)
&
$\begin{pmatrix}
(R,R) & (S,T) \\
(T,S) & (P,P)
\end{pmatrix}$

Action 0: cooperate; action 1: defect.
&
$T>P>S>R$.
&
Pure defection:
$\pi(0)=0,\ \pi(1)=1$.
\\
\midrule

Pure Coordination (PCo)
&
$\begin{pmatrix}
(a,a) & (b,b) \\
(b,b) & (a,a)
\end{pmatrix}$

Two equivalent coordination actions.
&
$a>b$.
&
Canonical mixed target:
$\pi(0)=\frac{1}{2},\ \pi(1)=\frac{1}{2}$.
\\
\midrule

Volunteer's Dilemma (Vol)
&
$\begin{pmatrix}
(b-c,b-c) & (b-c,b) \\
(b,b-c) & (0,0)
\end{pmatrix}$

Action 0: volunteer; action 1: defer.
&
$b>c>0$.
&
Mixed target:
$\pi(0)=1-\frac{c}{b}$,
$\pi(1)=\frac{c}{b}$.
\\
\midrule

Entry Game (Ent)
&
$\begin{pmatrix}
(d,d) & (a,b) \\
(b,a) & (c,c)
\end{pmatrix}$

Action 0: enter; action 1: stay out.
&
$a>c>d$ and $b>d$.
&
Mixed target:
$\pi(0)=\frac{a-c}{(a-c)+(b-d)}$,
$\pi(1)=1-\pi(0)$.
\\
\midrule

Battle of the Sexes (BoS)
&
$\begin{pmatrix}
(h,l) & (m,m) \\
(m,m) & (l,h)
\end{pmatrix}$

Asymmetric coordination game.
&
$h>l>m$.
&
Role-dependent mixed target:

Row:
$\pi_{\mathrm{row}}(0)=\frac{h-m}{h+l-2m}$.

Column:
$\pi_{\mathrm{col}}(0)=\frac{l-m}{h+l-2m}$.
\\
\midrule

Matching Pennies (MP)
&
$\begin{pmatrix}
(a,-a) & (-b,b) \\
(-c,c) & (d,-d)
\end{pmatrix}$

Generalized zero-sum game.
&
$a,b,c,d>0$.
&
Role-dependent mixed target:

Row:
$\pi_{\mathrm{row}}(0)=\frac{c+d}{a+b+c+d}$.

Column:
$\pi_{\mathrm{col}}(0)=\frac{b+d}{a+b+c+d}$.
\\
\midrule

Trust Game (Tru)
&
$\begin{pmatrix}
(e+r,e+r) & (0,e+g) \\
(e,e) & (e,e)
\end{pmatrix}$

Row: trustor. Column: trustee.
&
$g>r>0$ and $e>0$.
&
Pure target:
$\pi(0)=0,\ \pi(1)=1$.

Trustor does not trust; trustee defects.
\\
\midrule

Inspection Game (Ins)
&
$\begin{pmatrix}
(b-c,w) & (b-c,w-p) \\
(b,w) & (0,w+g)
\end{pmatrix}$

Row: inspector. Column: worker.
&
$b>c>0$ and $w\geq p>g>0$.
&
Role-dependent mixed target:

Inspector:
$\pi_{\mathrm{inspector}}(0)=\frac{g}{g+p}$.

Worker:
$\pi_{\mathrm{worker}}(0)=\frac{b-c}{b}$.
\\
\midrule

Rock-Paper-Scissors (RPS)
&
$\begin{pmatrix}
(0,0) & (-1,1) & (1,-1) \\
(1,-1) & (0,0) & (-1,1) \\
(-1,1) & (1,-1) & (0,0)
\end{pmatrix}$

Actions 0--2: rock, paper, scissors.
&
Fixed zero-sum payoff matrix.
&
Uniform mixed target:
$\pi(0)=\pi(1)=\pi(2)=\frac{1}{3}$.
\\
\midrule

Bertrand Duopoly (Ber)
&
$\begin{pmatrix}
(R,R) & (S,T) & (-c,0) \\
(T,S) & (P,P) & (-c,0) \\
(0,-c) & (0,-c) & (0,0)
\end{pmatrix}$

Action 0: high price; action 1: low price; action 2: marginal cost.
&
$T>R>0>P>S$ and $c>0$.
&
Pure marginal-cost target:
$\pi(0)=0,\ \pi(1)=0,\ \pi(2)=1$.
\\

\bottomrule
\end{longtable}

\renewcommand{\arraystretch}{1}

%% file: references.bib
@inproceedings{achille2019task2vec,
  title={Task2Vec: Task Embedding for Meta-Learning},
  author={Alessandro Achille and Michael Lam and Rahul Tewari and Avinash Ravichandran and Subhransu Maji and Charless C. Fowlkes and Stefano Soatto and Pietro Perona},
  booktitle={Proceedings of the IEEE/CVF International Conference on Computer Vision (ICCV)},
  year={2019},
  note={arXiv:1902.03545}
}

@Article{Omidshafiei2020,
author={Omidshafiei, Shayegan
and Tuyls, Karl
and Czarnecki, Wojciech M.
and Santos, Francisco C.
and Rowland, Mark
and Connor, Jerome
and Hennes, Daniel
and Muller, Paul
and P{\'e}rolat, Julien
and De Vylder, Bart
and Gruslys, Audrunas
and Munos, R{\'e}mi},
title={Navigating the landscape of multiplayer games},
journal={Nature Communications},
year={2020},
month={Nov},
day={05},
volume={11},
number={1},
pages={5603},
issn={2041-1723},
doi={10.1038/s41467-020-19244-4},
url={https://doi.org/10.1038/s41467-020-19244-4}
}

@inproceedings{zamir2018taskonomy,
  title={Taskonomy: Disentangling Task Transfer Learning},
  author={Amir R. Zamir and Alexander Sax and William Shen and Leonidas Guibas and Jitendra Malik and Silvio Savarese},
  booktitle={Proceedings of the IEEE Conference on Computer Vision and Pattern Recognition (CVPR)},
  year={2018},
  note={arXiv:1804.08328}
}

@inproceedings{vu2020transferability,
  title={Exploring and Predicting Transferability across NLP Tasks},
  author={Tu Vu and Tong Wang and Tsendsuren Munkhdalai and Alessandro Sordoni and Adam Trischler and Andrew Mattarella-Micke and Subhransu Maji and Mohit Iyyer},
  booktitle={Proceedings of the 2020 Conference on Empirical Methods in Natural Language Processing (EMNLP)},
  year={2020},
  note={arXiv:2005.00770}
}

@inproceedings{nguyen2020leep,
  title={LEEP: A New Measure to Evaluate Transferability of Learned Representations},
  author={Cuong V. Nguyen and Tal Hassner and Matthias Seeger and Cédric Archambeau},
  booktitle={Proceedings of the 37th International Conference on Machine Learning (ICML)},
  year={2020},
  note={arXiv:2002.12462}
}

@inproceedings{you2021logme,
  title={LogME: Practical Assessment of Pre-trained Models for Transfer Learning},
  author={Kaichao You and Yong Liu and Jianmin Wang and Mingsheng Long},
  booktitle={Proceedings of the 38th International Conference on Machine Learning (ICML)},
  year={2021},
  note={arXiv:2102.11005}
}

@inproceedings{alvarezmelis2020otdd,
  title={Geometric Dataset Distances via Optimal Transport},
  author={David Alvarez-Melis and Nicolò Fusi},
  booktitle={Advances in Neural Information Processing Systems (NeurIPS)},
  year={2020},
  note={arXiv:2002.02923}
}

@inproceedings{daroya2024task2box,
  title={Task2Box: Box Embeddings for Modeling Asymmetric Task Relationships},
  author={Rangel Daroya and Aaron Sun and Subhransu Maji},
  booktitle={Proceedings of the IEEE/CVF Conference on Computer Vision and Pattern Recognition (CVPR)},
  pages={28827--28837},
  year={2024},
  note={arXiv:2403.17173}
}

@inproceedings{lin2024consistency,
  title={Exploring the Effectiveness and Consistency of Task Selection in Intermediate-Task Transfer Learning},
  author={Pin-Jie Lin and Miaoran Zhang and Marius Mosbach and Dietrich Klakow},
  booktitle={Proceedings of the 62nd Annual Meeting of the Association for Computational Linguistics (Volume 4: Student Research Workshop)},
  pages={170--185},
  year={2024},
  note={arXiv:2407.16245}
}

@article{akata2023repeated,
  title={Playing repeated games with large language models},
  author={Elif Akata and Lion Schulz and Julian Coda-Forno and Seong Joon Oh and Matthias Bethge and Eric Schulz},
  journal={Nature Human Behaviour},
  volume={9},
  number={7},
  pages={1380--1390},
  year={2025},
  note={arXiv:2305.16867}
}

@inproceedings{fan2023rational,
  title={Can Large Language Models Serve as Rational Players in Game Theory? A Systematic Analysis},
  author={Caoyun Fan and Jindou Chen and Yaohui Jin and Hao He},
  booktitle={Proceedings of the AAAI Conference on Artificial Intelligence},
  volume={38},
  pages={17960--17967},
  year={2024},
  note={arXiv:2312.05488}
}

@article{gandhi2023strategic,
  title={Strategic Reasoning with Language Models},
  author={Kanishk Gandhi and Dorsa Sadigh and Noah D. Goodman},
  journal={arXiv preprint arXiv:2305.19165},
  year={2023},
  note={arXiv:2305.19165}
}

@inproceedings{duan2024gtbench,
  title={GTBench: Uncovering the Strategic Reasoning Limitations of LLMs via Game-Theoretic Evaluations},
  author={Jinhao Duan and Renming Zhang and James Diffenderfer and Bhavya Kailkhura and Lichao Sun and Elias Stengel-Eskin and Mohit Bansal and Tianlong Chen and Kaidi Xu},
  booktitle={Advances in Neural Information Processing Systems 37 (NeurIPS 2024)},
  year={2024},
  note={arXiv:2402.12348}
}

@article{gemp2024steering,
  title={Steering Language Models with Game-Theoretic Solvers},
  author={Ian Gemp and Roma Patel and Yoram Bachrach and Marc Lanctot and Vibhavari Dasagi and Luke Marris and Georgios Piliouras and Siqi Liu and Karl Tuyls},
  journal={arXiv preprint arXiv:2402.01704},
  year={2024},
  note={arXiv:2402.01704}
}

@inproceedings{kempinski2025gameofthoughts,
  title={Game of Thoughts: Iterative Reasoning in Game-Theoretic Domains with Large Language Models},
  author={Benjamin Kempinski and Ian Gemp and Kate Larson and Marc Lanctot and Yoram Bachrach and Tal Kachman},
  booktitle={Proceedings of the 24th International Conference on Autonomous Agents and Multiagent Systems (AAMAS)},
  pages={1088--1097},
  publisher={IFAAMAS},
  address={Richland, SC},
  year={2025}
}

@article{payne2025strategic,
  title={Strategic Intelligence in Large Language Models: Evidence from Evolutionary Game Theory},
  author={Kenneth Payne and Baptiste Alloui-Cros},
  journal={arXiv preprint arXiv:2507.02618},
  year={2025},
  note={arXiv:2507.02618}
}

@inproceedings{sun2025survey,
  title={Game Theory Meets Large Language Models: A Systematic Survey},
  author={Haoran Sun and Yusen Wu and Yukun Cheng and Xu Chu},
  booktitle={Proceedings of the Thirty-Fourth International Joint Conference on Artificial Intelligence (IJCAI)},
  pages={10669--10677},
  year={2025},
  note={arXiv:2502.09053}
}

@article{lore2024strategic,
  title={Large Model Strategic Thinking, Small Model Efficiency: Transferring Theory of Mind in Large Language Models},
  author={Nunzio Lorè and Sepehr Ilami and Babak Heydari},
  journal={arXiv preprint arXiv:2408.05241},
  year={2024},
  note={arXiv:2408.05241}
}

@article{marris2023equilibrium,
  title={Equilibrium-Invariant Embedding, Metric Space, and Fundamental Set of $2 \times 2$ Normal-Form Games},
  author={Luke Marris and Ian Gemp and Georgios Piliouras},
  journal={arXiv preprint arXiv:2304.09978},
  year={2023},
  note={arXiv:2304.09978}
}

@inproceedings{liu2024nfgtransformer,
  title={NfgTransformer: Equivariant Representation Learning for Normal-form Games},
  author={Siqi Liu and Luke Marris and Georgios Piliouras and Ian Gemp and Nicolas Heess},
  booktitle={Proceedings of the 12th International Conference on Learning Representations (ICLR)},
  year={2024},
  note={arXiv:2402.08393}
}

@article{wellman2024egta,
  title={Empirical Game-Theoretic Analysis: A Survey},
  author={Michael P. Wellman and Karl Tuyls and Amy Greenwald},
  journal={Journal of Artificial Intelligence Research},
  volume={82},
  pages={1017--1076},
  year={2025}
}

@article{omidshafiei2019alpharank,
  title={$\alpha$-Rank: Multi-Agent Evaluation by Evolution},
  author={Shayegan Omidshafiei and Christos Papadimitriou and Georgios Piliouras and Karl Tuyls and Mark Rowland and Jean-Baptiste Lespiau and Wojciech M. Czarnecki and Marc Lanctot and Julien Pérolat and Rémi Munos},
  journal={Scientific Reports},
  volume={9},
  number={1},
  pages={9937},
  year={2019},
  doi={10.1038/s41598-019-45619-9},
  note={arXiv:1903.01373}
}

@article{rapoport1966taxonomy,
  title={A Taxonomy of $2 \times 2$ Games},
  author={Rapoport, Anatol and Guyer, Melvin J.},
  journal={General Systems: Yearbook of the Society for General Systems Research},
  volume={11},
  pages={203--214},
  year={1966}
}

@book{robinson2005topology,
  title={The Topology of the $2 \times 2$ Games: A New Periodic Table},
  author={Robinson, David and Goforth, David},
  publisher={Routledge},
  year={2005}
}

@article{dafoe2020cooperative,
  title={Open Problems in Cooperative {AI}},
  author={Allan Dafoe and Edward Hughes and Yoram Bachrach and Tantum Collins and Kevin R. McKee and Joel Z. Leibo and Kate Larson and Thore Graepel},
  journal={arXiv preprint arXiv:2012.08630},
  year={2020},
  note={arXiv:2012.08630}
}

@inproceedings{park2023generative,
  title={Generative Agents: Interactive Simulacra of Human Behavior},
  author={Joon Sung Park and Joseph C. O'Brien and Carrie J. Cai and Meredith Ringel Morris and Percy Liang and Michael S. Bernstein},
  booktitle={Proceedings of the 36th Annual ACM Symposium on User Interface Software and Technology (UIST)},
  year={2023},
  note={arXiv:2304.03442}
}

@article{cerapalatsi2025cooperation,
  title={Large Language Models Replicate and Predict Human Cooperation Across Experiments in Game Theory},
  author={Cera Palatsi, Andrea and Martin-Gutierrez, Samuel and Cardenal, Ana S. and Pellert, Max},
  journal={arXiv preprint arXiv:2511.04500},
  year={2025},
  note={arXiv:2511.04500}
}

@article{kapoor2022leakage,
  title={Leakage and the Reproducibility Crisis in Machine-Learning-Based Science},
  author={Sayash Kapoor and Arvind Narayanan},
  journal={Patterns},
  volume={4},
  number={9},
  pages={100804},
  year={2023},
  note={arXiv:2207.07048},
  doi={10.1016/j.patter.2023.100804}
}

@inproceedings{agostinelli2022stable,
  title={How Stable Are Transferability Metrics Evaluations?},
  author={Andrea Agostinelli and Michal Pándy and Jasper Uijlings and Thomas Mensink and Vittorio Ferrari},
  booktitle={Proceedings of the European Conference on Computer Vision (ECCV)},
  pages={303--321},
  year={2022},
  note={arXiv:2204.01403}
}

@article{nash1951noncooperative,
  title={Non-Cooperative Games},
  author={Nash, John},
  journal={Annals of Mathematics},
  volume={54},
  number={2},
  pages={286--295},
  year={1951}
}

@book{camerer2003behavioral,
  title={Behavioral Game Theory: Experiments in Strategic Interaction},
  author={Camerer, Colin F.},
  publisher={Princeton University Press},
  year={2003}
}

@article{nakagawa2013r2,
  title={A General and Simple Method for Obtaining {R}$^2$ from Generalized Linear Mixed-Effects Models},
  author={Nakagawa, Shinichi and Schielzeth, Holger},
  journal={Methods in Ecology and Evolution},
  volume={4},
  number={2},
  pages={133--142},
  year={2013},
  doi={10.1111/j.2041-210x.2012.00261.x}
}

@article{shrout1979intraclass,
  title={Intraclass Correlations: Uses in Assessing Rater Reliability},
  author={Shrout, Patrick E. and Fleiss, Joseph L.},
  journal={Psychological Bulletin},
  volume={86},
  number={2},
  pages={420--428},
  year={1979}
}

@article{dafoe2021cooperative,
  title={Cooperative {AI}: Machines Must Learn to Find Common Ground},
  author={Allan Dafoe and Yoram Bachrach and Gillian Hadfield and Eric Horvitz and Kate Larson and Thore Graepel},
  journal={Nature},
  volume={593},
  number={7857},
  pages={33--36},
  year={2021}
}

@article{candogan2011flows,
  title={Flows and Decompositions of Games: Harmonic and Potential Games},
  author={Ozan Candogan and Ishai Menache and Asuman Ozdaglar and Pablo A. Parrilo},
  journal={Mathematics of Operations Research},
  volume={36},
  number={3},
  pages={474--503},
  year={2011}
}
